\documentclass[twocolumn,superscriptaddress,showkeys,preprintnumbers,amsmath,amssymb,prl]{revtex4}

\usepackage{graphicx,psfig,amssymb,color}
\usepackage{dcolumn}
\usepackage{bm}

\newcommand{\bfr}{{\bf r}}

\begin{document}


\title{Transition from Bose-Einstein condensate to Berezinskii-Kosterlitz-Thouless phase}
\author{T.~P. Simula}
\affiliation{Department of Physics, University of Otago, Dunedin, New Zealand}
\author{M.~D. Lee}
\affiliation{Clarendon Laboratory, Department of Physics, University of Oxford, Oxford, United Kingdom}
\author{D.~A.~W. Hutchinson}
\affiliation{Department of Physics, University of Otago, Dunedin, New Zealand}


\begin{abstract}
We obtain a phase diagram for a trapped two-dimensional ultra-cold Bose gas. We find a critical temperature above which the free energy of a state with a pair of vortices of opposite circulation is lower than the one for a vortex-free Bose-Einstein condensed ground state. We identify three distinct phases which are, in order of increasing temperature; a phase coherent Bose-Einstein condensate, a vortex pair plasma with a fluctuating condensate phase, and a thermal Bose gas. The existence of the vortex pair phase could be verified using current experimental setups.     
\end{abstract}


\maketitle

{\bf Introduction}

The theory of phase transitions is fundamental to the statistical mechanical description of nature \cite{LnL}. Technically, phase transitions involve non-analyticity in the free energy with a thermodynamic variable. The solid-liquid-gas transitions in water are common examples of first-order phase transitions, involving latent heat. In second-order phase transitions, such as the ferromagnetic transition, latent heat is absent at the critical point. If, in addition, the transition does not involve a symmetry breaking, it is classified as of infinite order. A famous example of such is the Berezinskii-Kosterlitz-Thouless (BKT) vortex unbinding transition in two-dimensional superfluid systems \cite{Berezinskii1971a,Kosterlitz1973a,Minnhagen1987a}. {Evidence for this phase transition has been experimentally demonstrated in liquid helium thin films \cite{Bishop1978a}, in superconducting Josephson-junction arrays \cite{Resnick1981a}, and in spin-polarized atomic hydrogen \cite{Safonov1998a}.}

Bose-Einstein condensation (BEC) is a purely quantum-statistical second-order phase transition occurring even in the absence of interactions \cite{Bose1924a,Einstein1924a,PitaevskiiBook}. Since the realization of BEC in 3D harmonic traps \cite{Anderson1995a,Davis1995a}, there has been an increased interest and debate on the nature of the normal-superfluid phase transition in the 2D Bose gas. It has long been established that BEC does not occur in homogeneous two-dimensional systems in the thermodynamic limit. This is formally proved in the Mermin-Wagner-Hohenberg theorem \cite{Mermin1966a,Hohenberg1967a}. Nevertheless, an external confining potential modifies the density of states in such a manner that, in a 2D ideal Bose gas, condensation into a single quantum state is predicted to occur at finite temperature \cite{Bagnato1991a}. It has been shown, however, that long-wavelength phase fluctuations may destroy the global phase coherence of the condensate below the ideal gas critical temperature, $T_0$, for BEC. The condensate can then only exist in the form of a so-called quasi-condensate \cite{Popov1983a,Petrov2000a,Prokofiev2002a,Prokofiev2002b,Andersen2002a,Al2002a,Gies2004a}. The connection between such a quasi-condensate and a BKT vortex-pair plasma remains unclear.
 \begin{figure}[!h]
\center
\includegraphics[viewport= 157 285 295 423,clip,width=86mm]{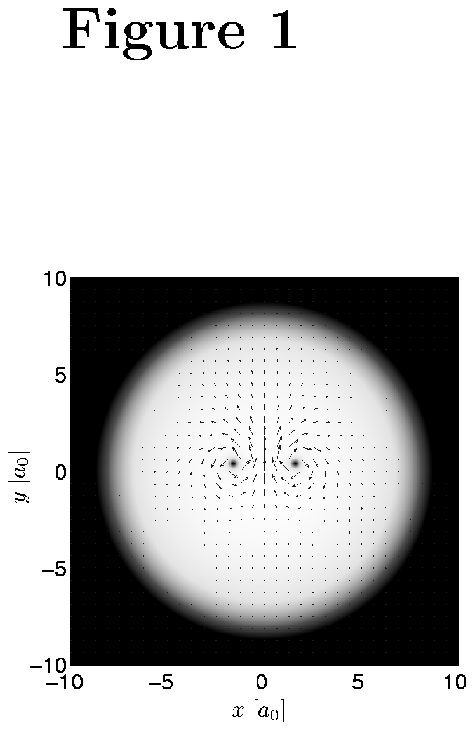}
\caption{Computed spatial distribution of the squared modulus of the order parameter containing a pair of vortices of opposite circulation.  The arrows represent the local superfluid currents. The vortex pair shown is widely separated for demonstrational purposes.}
\label{Fig1}
\end{figure}
{Suggested methods for achieving the required two-dimensionality for observing the BKT phase transition in dilute atomic gases include application of tight axial trapping \cite{Smith2005a}, use of optical lattice potentials \cite{Trombettoni2005a}, and rapidly rotating gases with hard-wall radial confinement \cite{Fischer2003a}.} In this Letter we show, based on a free energy argument, that a 2D harmonically trapped Bose gas may exhibit both the BEC and BKT-like superfluid phases, separated by what we believe to be a first-order phase transition. At higher temperature, the latter is expected to undergo a vortex unbinding transition to a normal state analogous to the usual BKT transition for the homogeneous system.  

{\bf Free energy}

We consider a dilute gas consisting of $N$ bosonic particles of mass $m$ spatially confined in two dimensions by an external harmonic potential $V(\bfr)=m\omega_\perp^2r^2/2$, where $\omega_\perp$ is the radial trap frequency. We assume the existence of a macroscopic ground state condensate wavefunction $\psi(\bfr)$, which also serves to describe the order parameter of the system in the BEC phase. Our aim is to calculate and compare the Helmholtz free energy $F=E-TS$ for a phase coherent ground state and for an excited state containing a vortex pair in the ordering field, see Fig.~\ref{Fig1}, for various configurations. Due to entropy considerations, one expects to find a critical temperature, $T_c$, above which a thermally activated transition to the state containing vortex-pair excitations becomes favourable.

Isolating the contribution due to the condensate, $E_0(T)$, the total internal energy of the system may be written as $E(T)= E_0(T)+\tilde{E}(T)$. The required condensate energy $E_0(T)$ is determined by the functional
\begin{equation}
E_0(T)=\int \left(\frac{\hbar^2}{2m}|\nabla\psi(\bfr)|^2 +V(\bfr)|\psi(\bfr)|^2 + \frac{g}{2}|\psi(\bfr)|^4\right) {\rm d}\bfr
\label{EGY}
\end{equation}
where $g$ is the constant coupling parameter for the particle interactions, and $N_0=\int |\psi(\bfr)|^2 {\rm d}\bfr$ denotes the number of particles in the condensate. In order to implicitly incorporate the temperature dependence in $E_0(T)$, we employ the 2D ideal-gas result for the condensate fraction
\begin{equation}
\frac{N_0}{N}=1-\left(\frac{T}{T_0}\right)^2
\end{equation}Alternatively, the same 
where $k_BT_0=\hbar\omega_\perp\sqrt{N/\zeta(2)}$ and $\zeta$ is the Riemann zeta function. The shift in the critical temperature from the 2D ideal-gas prediction {\bf $T_0$} is only of a few percent, due to the weak particle interactions \cite{Gies2004a}. We therefore expect the above approximation to yield a good estimate for the pair creation transition temperature in the system considered here. The internal energies for topologically distinct order parameter configurations are calculated {both analytically within the hydrodynamic approximation and} computationally as described along with the numerical methods. 

To evaluate the free energy difference, one also needs to calculate the entropy change $\Delta S=k_B\ln W$, due to the multiplicity of order parameter configurations containing a pair of vortices. The statistical weight, $W=2\pi R_{TF}^2/\xi^2$, is obtained by allowing one vortex to reside anywhere within the Thomas-Fermi radius $R_{TF}=\sqrt{2\mu/m\omega^2_\perp}$ of the condensate, where $\mu$ is its chemical potential. The partner vortex then has $2\pi$ available nearest neighbour sites, assuming that the pair is closely bound when it forms. Approximating the radius of the area occupied by a vortex to be equal to the healing length, $\xi=\sqrt{\hbar^2/2m\mu}$, of the order parameter, we obtain the configurational entropy $\Delta S(T)=k_B\ln{(8\pi\mu^2/\hbar^2\omega_\perp^2)}$ due to the available microstates containing a pair of vortices. Here we assume that all of these microstates possess the same internal energy although in a strict sense this is not true, since the gas is inhomogeneous. {However, microstates of equal energy with the vortex pair residing outside the condensate centre may be chosen to be ones with correspondingly larger vortex pair separation, balancing the energy reduction due to the inhomogeneous superfluid density}. By applying the Thomas-Fermi approximation for the chemical potential, the relevant free energy difference per particle is given by the formula 
\begin{equation}
\frac{\Delta F(T)}{N}=\frac{\Delta E_0(T)}{N_0}\frac{N_0}{N} - \frac{T}{T_0}\frac{\ln{(8gN_0m/\hbar^2)}}{\sqrt{\zeta(2)N}} \hbar\omega_\perp.
\label{deltaf}
\end{equation}
Guided by the results for the excitation energy of a single vortex at finite temperatures \cite{Simula2002a}, we have assumed that the energy difference associated with uncondensed atoms $\Delta\tilde{E}(T)\ll \Delta E_0(T)$ in Eq.~(\ref{deltaf}). In the same fashion, the entropy difference due to the quasiparticle excitations is neglected. 

The critical temperature, $T_c$, for the pair creation transition is determined by the condition $\Delta F(T_c)=0$. Letting $\bfr^+$ and $\bfr^-$ denote the positions of the oppositely circulating vortices, we define the energy of a pair excitation per condensate particle $\alpha(\bfr^+,\bfr^-)=\Delta E_0(T)/N_0$. Hence, the transition temperature is implicitly given by the formula
\begin{equation}
\frac{T_c}{T_0}=\frac{\alpha}{\hbar\omega_\perp}\frac{N_0}{\sqrt{N}}\frac{\sqrt{\zeta(2)}}{\ln{(8gN_0m/\hbar^2)}}.
\end{equation}

{In the hydrodynamic approximation the energy of a vortex pair may be estimated to be \cite{PitaevskiiBook}
\begin{equation}
E_{vp}=\frac{2\pi n_0\hbar^2}{m}\ln{\frac{D}{\xi}}+2E_c,
\end{equation}
where $n_0$ is the superfluid particle density, $D=2\xi$ is the pair separation, and $E_c$ denotes the energy associated with a single vortex core. Assuming that the condensate density within an isolated vortex core increases quadratically as a function of the distance from its centre, we obtain
\begin{equation}
E_{vp}=\frac{2\pi\mu\hbar^2}{gm}\left(\ln{2}+\frac{1}{2}\right)
\label{vegy}
\end{equation}
for a vortex pair located in the vicinity of the trap centre. Inserting Eq.~(\ref{vegy}) for $\Delta E_0(T)$ in $\alpha$, one finds that the critical temperature is given by 
\begin{equation}
\frac{T_c}{T_0}=\sqrt{\frac{\beta}{\beta+g\ln^2(8gmN_0/\hbar^2)}}.
\end{equation}
where, $\beta=2\pi^3\hbar^2(\ln 2+1/2)^2/3m$. In the limit of the non-interacting ideal-gas $\lim_{g\to 0}(T_c/T_0)=1$ and the BKT phase vanishes. In homogeneous systems the BEC phase does not exist---a result that can be recovered by considering the large $g$ limit in which case the condensate density becomes homogeneous on the length scale of the vortex core. This conclusion is also reached if the thermodynamic limit, $\omega\to0$ and $N\to\infty$ with $N\omega^2$ held constant, is taken \cite{PitaevskiiBook}. 
}

\begin{figure}[!h]
\center
\includegraphics[viewport= 170 400 420 610,clip,width=86mm]{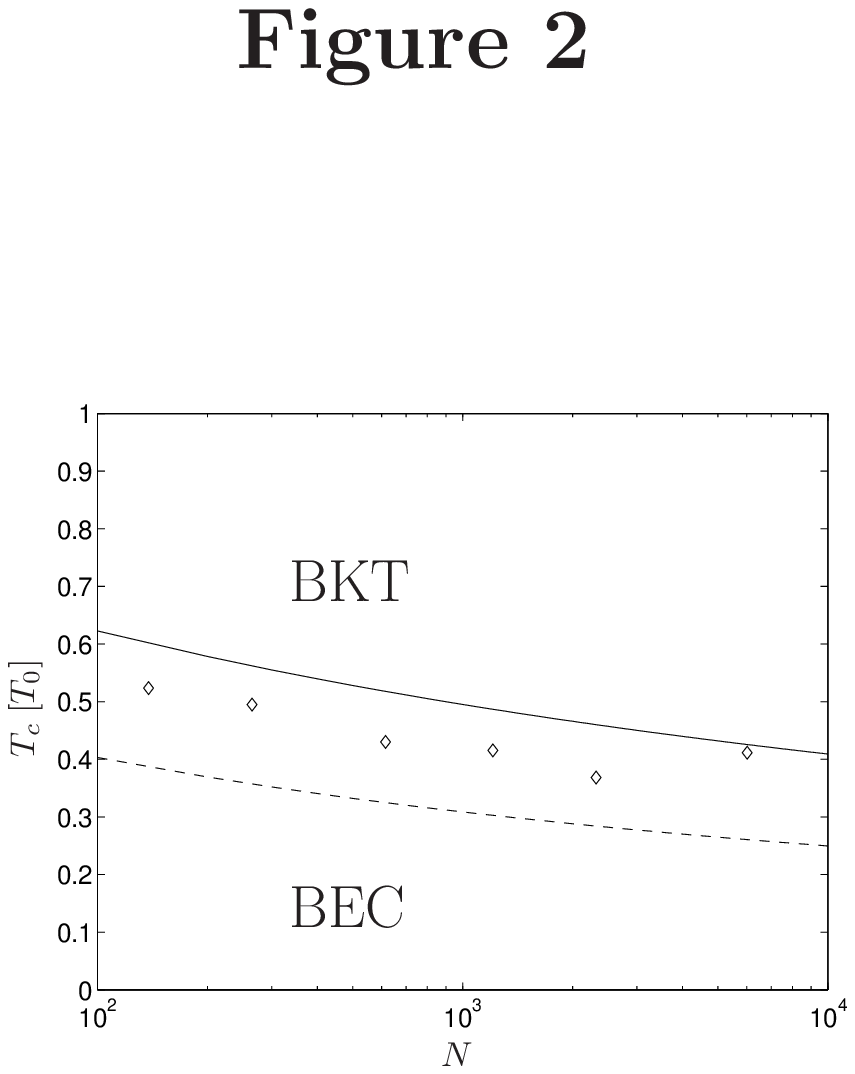}
\caption{{Solution of Eq.~(\ref{TEECEE}) as a function of the total particle number $N$. The phase boundaries are found by applying different estimates for the vortex pair energy $\alpha$. The diamonds correspond to the full numerical calculation whereas the continuous lines are plotted using Eq.~(\ref{vegy}) with (solid line), or without (dashed line) the core energy contribution.}}
\label{Fig2}
\end{figure}

{\bf Numerical results}

{In order to assess the validity of the hydrodynamic approximation we calculate the pair excitation energy $\alpha$ by solving numerically the Gross-Pitaevskii equation} 
\begin{equation}
\left(\frac{\hbar^2}{2m}\nabla^2  +V(\bfr)  + g|\psi(\bfr)|^2\right)\psi(\bfr)=\mu\psi(\bfr)
\label{GPE}
\end{equation}
for a variety of order parameter configurations. The vortex pair states are generated by initially phase imprinting the two vortices with opposite circulation in the order parameter field as depicted in Fig.~\ref{Fig1}. The subsequent damped real-time propagation of $\psi(\bfr)$ under the Hamiltonian corresponding to the lhs of Eq.~(\ref{GPE}) {serves to provide an ensemble of variational wavefunctions whose energies are given by the energy functional, Eq.~(\ref{EGY}).} 

Computationally, all dimensional parameters in Eq.~(\ref{GPE}) can be embedded into a single constant $C=gN_0/\hbar\omega_\perp a_0^2$ where $a_0=\sqrt{\hbar/2m\omega_\perp}$ defines the characteristic length scale in the system. We model the experimentally achievable quasi-2D systems and hence the coupling constant $g=\sqrt{8\pi}\hbar^2a/ma_z$, where $a$ is the $s$-wave scattering length, $a_z=\sqrt{\hbar/m\omega_z}$, and $\omega_z$ is the axial trapping frequency. We have also verified that using this quasi-2D value for the particle interactions is indistinguishable from the full T-matrix result \cite{Petrov2000a,Lee2002a}. The equation for $C$ may thus be expressed as 
\begin{equation}
C=4\sqrt{2\pi}N\frac{a}{a_z}\left(1-\left(\frac{T}{T_0}\right)^2\right)
\end{equation}
and therefore, by altering the single parameter $C$, one can explore the full parameter space of the problem. Thus, once the vortex pair excitation energy $\alpha$ is calculated, the critical temperature is obtained from the equation
\begin{equation}
\frac{T_c}{T_0}=\frac{\alpha}{\hbar\omega_\perp}\frac{N_0}{\sqrt{N}}\frac{\sqrt{\zeta(2)}}{\ln{(16\sqrt{2\pi}N_0a/a_z)}}.
\label{TEECEE}
\end{equation}

Modelling the 2D experimental system at Oxford \cite{Smith2005a}, we choose $\{\omega_\perp,\omega_z\}=2\pi\times\{3,2200\}$ Hz. We have calculated the pair excitation energy $\alpha$ for a range of parameters $C$, $\bfr^+$ and $\bfr^-$. This energy depends on the separation $|\bfr^+-\bfr^-|$ of the vortices due to their logarithmic attractive interaction {and is proportional to the local condensate density at the vortex locations. The main difficulty in obtaining accurate estimates for the energies of the vortex-pair states is the fact that well defined vortex-pair eigenstates do not exist. Additional problems arise with large condensates for which the vortex cores are very small and the energy of the ground state is many orders of magnitude larger than that of the vortex pair excitation.}

{The solution of Eq.~(\ref{TEECEE}) is plotted in Fig.~(\ref{Fig2}) as a function of the total particle number $N$, using the Oxford trap parameters and three different estimates for the vortex pair energy $\alpha$. The diamonds correspond to the full numerical calculation whereas the continuous lines are plotted using Eq.~(\ref{vegy}) with (solid line), or without (dashed line) the core energy contribution.

The free energy difference changes sign at the phase boundary and vortex pair creation becomes favourable at higher temperatures. It is to be noted that in a real system the thermally activated vortex pair nucleation would be likely to occur first near the low density perimeter of the condensate and in this respect the estimated phase boundaries in Fig.~(\ref{Fig2}) correspond to upper bounds on the true critical temperature.}

\begin{figure}[!h]
\center
\includegraphics[viewport= 175 410 370 562,clip,width=86mm]{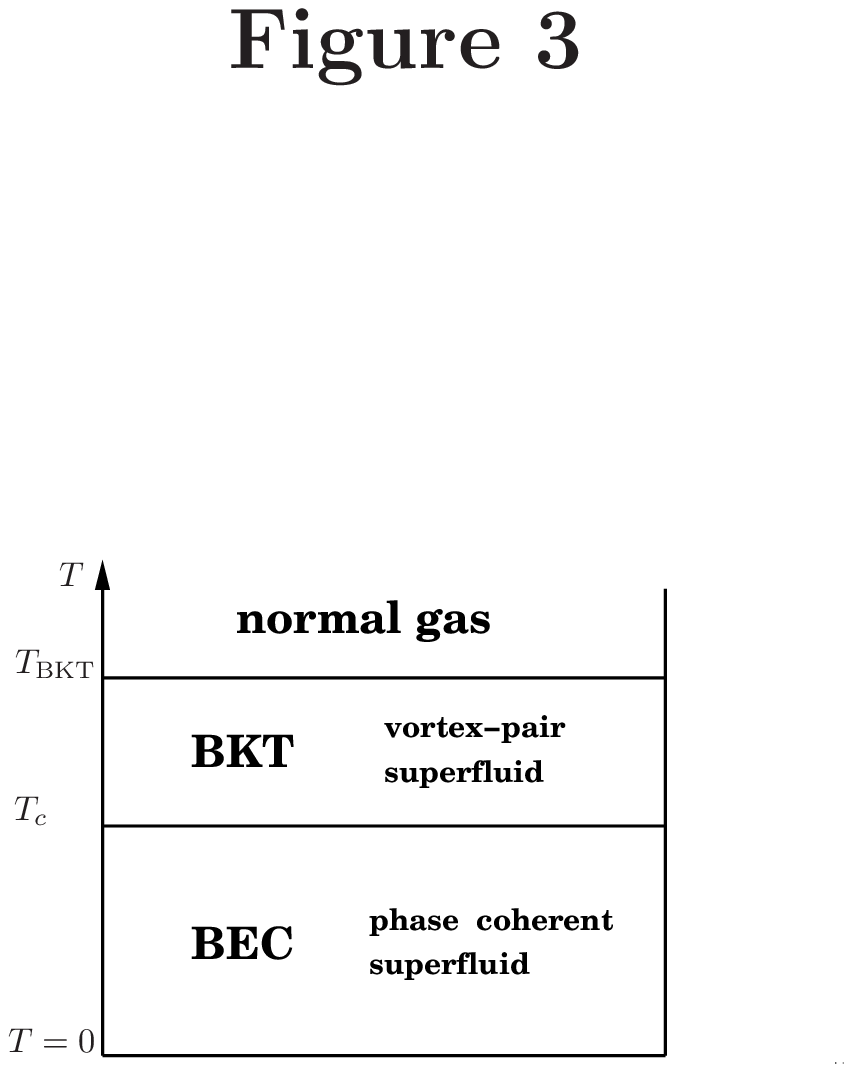}
\caption{Phase diagram for a 2D trapped Bose gas at ultra-low temperatures. }
\label{Fig3}
\end{figure}

Thus we arrive at the phase diagram shown in Fig.~(\ref{Fig3}) for the dilute 2D gas of bosonic particles at ultra-low temperatures. For the lowest temperatures the gas exists in a Bose-Einstein condensed state. With increasing temperature the phase coherence is lost due to the long-wavelength phase fluctuations and the gas may undergo an entropy driven, first-order phase transition to the Berezinskii-Kosterlitz-Thouless-like phase. This phase consists of bound pairs of vortices with opposite circulation. The gas exhibits superfluidity on both sides of this pair creation transition, although above it, only a topological order exists in terms of correlations between the locations of the vortices. These correlations yield algebraic long-range order, allowing the system to remain in a superfluid state. At higher temperatures the paired vortices unbind at the BKT superfluid to normal phase transition. Our present formalism does not allow us to probe this upper transition in great detail. {Nevertheless, the binding energy of a pair is roughly equal to the pair creation energy, as is verified by our numerical simulations. Therefore, the temperature at which single isolated vortices can form spontaneously would be twice the temperature at which bound vortex pairs are created, suggesting the temperature range for the BKT phase to be roughly equal to that of the phase coherent BEC phase. This estimation is also in reasonable agreement with the ideal gas prediction for the superfluid-normal transition temperature $T_0$. At any rate, the width of the BKT phase should exceed the temperature resolution in currently feasible experiments. In this context it is worth re-emphasizing that in a homogeneous system, BEC only occurs at $T=0$ and therefore the BKT phase is the only superfluid phase of the system at finite temperature. However, in the presence of an external confining potential, phase fluctuations are suppressed, allowing the formation of a phase coherent BEC. This leads to the existence of two distinct superfluid phases.}

{\bf Discussion}

We have calculated the free energy for a dilute harmonically trapped two-dimensional Bose gas at ultra-low temperatures. We have shown that in such a system there may exist two different superfluid phases---a phase coherent Bose-Einstein condensate and a phase fluctuating Berezinskii-Kosterlitz-Thouless-like phase. We obtain an estimate for the critical temperature separating these two phases in terms of the external system parameters. Since we are interested in the onset temperature for spontaneous pair formation, it suffices to only consider a single pair of vortices instead of the full interacting pair vortex plasma. After this an avalanche of pair creation may occur. Unbinding of these vortex pairs eventually leads to the destruction of superfluidity in the system. 

The approach adopted here could be extended to take the thermal quasiparticle effects explicitly into account, but these effects are of higher order and are not expected to alter the qualitative picture obtained. This is especially true since we are interested in the free energy difference rather than its absolute value. In fact, the most significant effect neglected in this respect is probably the finite-temperature core broadening of vortices due to the thermal cloud pressure, which is only significant at high temperatures with very small condensate fraction. The transition to the BKT phase is predicted to occur at much lower temperatures, however, validating our approach. The actual transition temperature may also be lower than our prediction, since we consider vortex pair creation in the vicinity of the trap centre where the self-energy of the pair is the highest. In practice one expects the spontaneous pair creation to begin near the low density boundaries of the condensate, {which also opens the possibility of the co-existence of the BEC and BKT phases.} 

The concept of the quasi-condensate is inherent within the BKT phase, with the vortex pairs located at the boundaries of the locally phase coherent domains. It is possible that, at lower temperatures, phase fluctuations may already destroy the global phase coherence, yielding a quasi-condensate prior to the formation of the BKT phase. Thus the pair-creation transition implies a quasi-condensate but not necessarily the converse.

We suggest that it would be possible to directly observe the paired vortices in the BKT-like phase. The BKT phase may pose observational difficulties even if it were present in the system. However, one could possibly observe the paired vortices using standard imaging procedures after a ballistic expansion of the condensate. Another, theoretically tempting possibility, could be to separate the vortices of opposite circulation by breaking the time-reversal symmetry. This could be done by rotating the condensate below the rotationally activated vortex nucleation frequency. Subsequently one could measure the residual angular momentum of the system, whose finite value would be a clear indication of the pre-existence of vortex pairs in the BKT phase, since the direct formation of single vortices is forbidden.

\begin{acknowledgments}
We would like to thank Chris Foot, and Keith Burnett for useful discussions and Nathan Smith for pointing out the importance of the variant vortex pair self-energy. Financial support from the Academy of Finland, the Marsden Fund of the Royal Society of New Zealand, and the Royal Society (London) is acknowledged. 
\end{acknowledgments}

\end{document}